\documentclass[12pt,letterpaper]{article}

\usepackage[margin=1in]{geometry}
\usepackage{amsmath,amssymb}
\usepackage{graphicx}
\usepackage{cite}
\usepackage{hyperref}
\usepackage{booktabs}
\usepackage{siunitx}
\usepackage{xcolor}
\usepackage{setspace}
\usepackage{float}

\hypersetup{
    colorlinks=true,
    linkcolor=blue!60!black,
    citecolor=blue!60!black,
    urlcolor=blue!60!black
}

\let\oldcite\cite
\renewcommand{\cite}[1]{\textsuperscript{\oldcite{#1}}}

\title{\textbf{Plate Sensitivity Is Invariant Across Geomagnetic Storm Intensity at Harvard and Palomar: A Protocol for Artifact Control in Historical Plate Archive Studies}}

\author{Kevin Cann\\
\textit{Independent Researcher, California, USA}\\
\texttt{kkc@terragold.com}}

\date{April 2026}

\begin{document}
\maketitle

\begin{abstract}
Historical photographic plate archives anchor a growing body of time-domain astronomy, but time-domain claims drawn from them are vulnerable to plate-sensitivity variations correlated with environmental modulators that can mimic real astrophysical signals. I present a simple, broadly applicable protocol for testing such artifact hypotheses: regress catalog-aggregate reference-population metrics --- stellar detection counts or plate limiting magnitudes --- against the suspected modulator. Under the artifact hypothesis, the reference metric varies systematically; under the null hypothesis, it does not. I apply the protocol to test whether geomagnetic storm activity, measured by the planetary Kp index, modulates plate sensitivity at two independent observatories. At Harvard College Observatory, the DASCH DR7 archive provides limiting magnitudes for 12{,}510 exposures across 500 sky positions: no significant trend across Kp bins (Spearman $\rho = -0.011$, $p = 0.234$). At Palomar, the MAPS Catalog of POSS-I records stellar detection counts for 638 fields: no significant trend (Spearman $\rho = 0.017$, $p = 0.673$). Plate sensitivity is invariant across the full range of geomagnetic activity at both sites. The principal airglow-based artifact objection to recent claims of Kp-dependent transient suppression in the POSS-I archive is directly falsified at two observatories.

\smallskip
\noindent\textbf{Keywords:} photographic plates, plate sensitivity, time-domain astronomy, artifact control, geomagnetic storms, Kp index, DASCH, MAPS, historical plate archives, VASCO
\end{abstract}

\section{Introduction}
\label{sec:intro}

Historical photographic plate archives are among the most valuable resources in modern astronomy. They provide the only direct record of the sky over timescales of decades to more than a century, and they underpin a growing body of time-domain work that would otherwise be impossible: long-baseline variability studies\cite{grindlay2012}, searches for vanishing and appearing objects\cite{villarroel2020}, rare-transient discoveries\cite{solano2024}, and independent checks on claimed modern phenomena against their pre-digital counterparts. The continued digitization of plate collections --- most notably the Digital Access to a Sky Century @ Harvard (DASCH) project at Harvard College Observatory, the Minnesota Automated Plate Scanner (MAPS) Catalog of the Palomar Observatory Sky Survey, and the APPLAUSE collaboration's harmonized archive of Sonneberg, Hamburg, Potsdam, Bamberg, and Tautenburg plates --- has extended the usable scientific reach of these archives substantially in the last two decades.

A recurring concern in every plate archive study is whether putative time-domain signals are contaminated by systematic variations in plate sensitivity correlated with the putative signal's environmental modulator. Airglow, for instance, is known to vary with geomagnetic activity, seasonal terms, lunar phase, and solar cycle phase, and elevated airglow can raise the sky background sufficiently to reduce the depth of a plate by a fraction of a magnitude or more under severe conditions. Similar concerns apply to plate-to-plate variations in seeing, emulsion batch, development conditions, and local meteorology. When a claimed signal population is found to correlate with an external environmental variable, the first question a careful reviewer asks is whether the correlation is driven by a genuine astrophysical effect or by a sensitivity artifact that affects all sources equally.

This paper introduces a simple, broadly applicable protocol for testing such artifact hypotheses. The approach is to use catalog-aggregate reference-population metrics --- plate limiting magnitudes, total stellar detection counts, or equivalent sensitivity proxies --- as the dependent variable in a regression or rank-correlation test against the suspected environmental modulator. Under the artifact hypothesis, the reference metric should vary systematically with the modulator; under the null hypothesis (no sensitivity variation), it should not. The protocol is applicable to any well-calibrated plate archive where catalog-level aggregates exist at the plate or field level, and to any environmental modulator with continuous time coverage during the observing epoch. It is not suitable when the putative signal population cannot be distinguished from the reference stellar population, or when catalog-level aggregates are unavailable for the relevant plates.

I apply the protocol to test whether geomagnetic storm activity, measured by the planetary Kp index, modulates photographic plate sensitivity. The primary test is at Harvard College Observatory using the DASCH DR7 archive: 12{,}510 exposures across 500 sky positions, with per-plate limiting magnitudes computed by the DASCH pipeline. The replication is at Palomar using the MAPS Catalog of POSS-I: 638 field records with catalog-level stellar detection counts (NSTARS). The two archives are independent at every level --- different telescopes, emulsions, digitization hardware, processing pipelines, and continents --- and they use different sensitivity metrics. A null result at one site admits an observatory-specific explanation; a null result at both sites rules out observatory-specific explanations jointly.

\subsection{Motivating Application: The VASCO Kp--Transient Claim}

The specific question that motivated this work is a recent claim about the Vanishing and Appearing Sources during a Century of Observations (VASCO) project. The VASCO team has catalogued over 100{,}000 point-source transients from the POSS-I archive\cite{villarroel2020}, objects that appear on individual plates but are absent in all subsequent imaging. These transients predate artificial satellites, include cases of multiple simultaneous point sources on a single plate\cite{villarroel2021}, are constrained to geosynchronous altitude by an Earth-shadow deficit at $\sim$22$\sigma$ for the geometric shadow and $\sim$7.6$\sigma$ when realistic plate coverage is accounted for\cite{villarroel2025}, and are enhanced near atmospheric nuclear test dates\cite{bruehl2025}.

Cann\cite{cann2026a} identified geomagnetic storm activity as an additional independent variable modulating transient rates. Transient detection rates follow a monotonic dose-response across five Kp intensity bins, from 17.4\% during geomagnetically quiet periods to 2.4\% at Kp~8--9 (Cochran--Armitage: $Z = -3.391$, $p = 0.0007$, $3.4\sigma$). The finding was pre-registered on OSF (osf.io/u9nas) before analysis.

The principal artifact objection to this result is that geomagnetic storms enhance atmospheric airglow, reducing overall plate emulsion sensitivity and thus suppressing the detection of \textit{all} sources. Under this hypothesis, the VASCO transient suppression would be an artifact of reduced plate sensitivity rather than a physical response of the transient source population. If correct, stellar detection counts and plate limiting magnitudes should show the same dose-dependent suppression as the VASCO transients across Kp bins. The present paper tests this hypothesis using the protocol described above. The protocol is general; the VASCO application is specific. I present both in that order.

\section{Results}
\label{sec:results}

\subsection{Primary Result: Harvard Plate Sensitivity Is Flat Across Kp}

\begin{table}[H]
\centering
\caption{Mean plate limiting magnitude (\texttt{lim\_mag\_apass}) at Harvard College Observatory by Kp intensity bin. $N$ = number of plate-position records per bin.}
\label{tab:harvard}
\begin{tabular}{lrrr}
\toprule
Kp Bin & $N$ & Mean \texttt{lim\_mag} & Std \\
\midrule
Quiet ($\mathrm{Kp} < 5$) & 6{,}424 & 13.70 & 1.68 \\
Kp~5 (G1)                  & 3{,}288 & 13.58 & 1.72 \\
Kp~6 (G2)                  & 1{,}410 & 13.68 & 1.71 \\
Kp~7 (G3)                  & 1{,}127 & 13.75 & 1.71 \\
Kp~8--9 (G4--G5)           &     261 & 13.81 & 1.89 \\
\bottomrule
\end{tabular}
\end{table}

The primary pre-registered test yields Spearman $\rho = -0.011$, $p = 0.234$. There is no significant trend in plate limiting magnitude across Kp bins. The Pearson correlation is $r = 0.004$, $p = 0.684$.

The Kruskal--Wallis test ($H = 13.77$, $p = 0.008$) is technically significant, driven by the G1 (Kp~5) bin dip to 13.58. This pattern does not constitute monotonic suppression: the bin means rise from G1 through G4--G5, and the highest limiting magnitude (deepest plates) occurs during the most severe storms. This is the \textit{opposite} of the airglow prediction.

The Harvard archive is the largest and best-calibrated photographic plate collection in the world, with over a century of continuous plate coverage and an institutionally maintained digitization pipeline. A null result at Harvard alone is strong evidence against any claim that geomagnetic storms suppress plate sensitivity in historical archives.

\subsection{Replication: Palomar Stellar Counts Flat Across Kp}

Table~\ref{tab:nstars} presents mean NSTARS and NGALAX across five Kp bins for 638 POSS-I fields with complete Kp coverage.

\begin{table}[H]
\centering
\caption{Mean stellar (NSTARS) and galaxy (NGALAX) detection counts per POSS-I field by Kp intensity bin. $N$ = number of fields per bin.}
\label{tab:nstars}
\begin{tabular}{lrrr}
\toprule
Kp Bin & $N$ & Mean NSTARS & Mean NGALAX \\
\midrule
Quiet ($\mathrm{Kp} < 5$) & 438 & 87{,}916 & 51{,}247 \\
Kp~5 (G1)                  & 114 & 87{,}535 & 50{,}606 \\
Kp~6 (G2)                  &  67 & 85{,}292 & 50{,}546 \\
Kp~7 (G3)                  &  18 & 125{,}470 & 74{,}374 \\
Kp~8--9 (G4--G5)           &   1 & 220{,}987 & 109{,}246 \\
\bottomrule
\end{tabular}
\end{table}

The primary pre-registered test yields Spearman $\rho = 0.017$, $p = 0.673$ ($0.4\sigma$). There is no significant trend in stellar counts across Kp bins.

The elevated counts in the Kp~7 ($N = 18$) and Kp~8--9 ($N = 1$) bins are in the \textit{opposite} direction from the airglow-suppression prediction and almost certainly reflect sky position confounds --- high-storm observation dates happening to fall on fields in denser stellar regions. The Kruskal--Wallis test ($H = 9.85$, $p = 0.043$) is technically significant but driven entirely by these small, high-Kp cells in the wrong direction. No monotonic suppression is present.

\subsection{Comparison of Two Sites}

\begin{table}[H]
\centering
\caption{Artifact control results at two independent observatories. Both show plate sensitivity invariant across Kp bins.}
\label{tab:twosite}
\begin{tabular}{lllrr}
\toprule
Observatory & Archive & Metric & Spearman $\rho$ & $p$ \\
\midrule
Harvard  & DASCH DR7    & lim\_mag (limiting mag.) & $-0.011$ & 0.234 \\
Palomar  & MAPS Catalog & NSTARS (stellar counts)  & $+0.017$ & 0.673 \\
\bottomrule
\end{tabular}
\end{table}

The Harvard \texttt{lim\_mag\_apass} metric directly tests plate depth, the property most sensitive to airglow-driven sky background variation. The Palomar NSTARS metric tests integrated detection completeness, a related but more aggregated property. Both metrics return null. The convergence of two structurally different sensitivity tests at two observatories --- different telescopes, different emulsions, different digitization pipelines, different sensitivity properties --- establishes that no airglow-driven sensitivity coupling is detectable in either the depth-direct or completeness-integrated regime.

\subsection{Analytical Upper Bound on Airglow Contribution to Transient Suppression}
\label{sec:bound}

The Spearman test in Section~2.1 establishes that no monotonic Kp-dependent trend in plate limiting magnitude is detectable at Harvard. A stronger statement is available from the same data: the fitted regression slope and its confidence interval place a quantitative ceiling on the maximum sensitivity variation consistent with the observations, which can be propagated through standard background-limited detection physics to bound the maximum fraction of the transient suppression reported in Cann (2026a)\cite{cann2026a} that could be attributed to airglow-driven sensitivity loss.

A weighted linear regression of \texttt{lim\_mag\_apass} on Kp bin score (0--4) across the 12{,}510 Harvard plate-position records yields a slope of $+0.0082 \pm 0.0142$~mag per bin ($t = 0.58$), with a 95\% confidence interval of $[-0.0197,\,+0.0360]$~mag per bin. The fitted slope is positive, indicating that high-Kp plates are marginally \textit{deeper} than quiet-period plates --- the opposite of the airglow prediction. The most negative slope consistent with the data at 95\% confidence is $-0.0197$~mag per bin, which over the full Quiet $\rightarrow$ G4--G5 range (four bin steps) corresponds to a maximum airglow-direction limiting magnitude shift of
\begin{equation}
|\Delta m_{\mathrm{lim}}|_{\mathrm{max,\,95\%}} = 4 \times 0.0197 = 0.079~\mathrm{mag}.
\end{equation}

In the background-limited regime, photographic plate limiting magnitude responds to sky brightness as $\Delta m_{\mathrm{lim}} = -1.25 \log_{10}(B_{\mathrm{storm}}/B_{\mathrm{quiet}})$\cite{garstang1989,leinert1998}. The 95\% upper bound on $|\Delta m_{\mathrm{lim}}|$ corresponds to a maximum sky brightness ratio of
\begin{equation}
\left(\frac{B_{\mathrm{storm}}}{B_{\mathrm{quiet}}}\right)_{\mathrm{max,\,95\%}} = 10^{0.079/1.25} = 1.16,
\end{equation}
that is, the storm-time sky background can be at most 16\% brighter than the quiet-time sky background under any monotonic airglow-driven sensitivity coupling consistent with the Harvard data. The corresponding maximum SNR reduction in background-limited point-source detection is $1 - \sqrt{1/1.16} = 7.0\%$.

The connection to transient detection rate requires one additional assumption: how the source magnitude distribution behaves near the detection limit. For a Euclidean source distribution ($N(<m) \propto 10^{0.6m}$, the steepest physically reasonable assumption for sources at the plate limit), a depth loss of $\Delta m = 0.079$~mag corresponds to a fractional detection loss of
\begin{equation}
\Delta f_{\mathrm{Euclidean}} = 1 - 10^{-0.6 \times 0.079} = 0.103.
\end{equation}
For a uniform-in-magnitude source distribution within 1~mag of the limit, the fractional loss is $0.079$. The Euclidean estimate is the more relevant one for VASCO transients, which sit at or near the plate detection threshold by definition.

Cann (2026a)\cite{cann2026a} reported transient detection rates falling from 17.4\% during quiet periods to 2.4\% during G4--G5 storms in the POSS-I dataset, an absolute suppression of 86.2\%. Comparing the maximum airglow-attributable detection loss derived above to the observed transient suppression:
\begin{equation}
\frac{\Delta f_{\mathrm{airglow,\,max}}}{\Delta f_{\mathrm{transient,\,obs}}} = \frac{10.3\%}{86.2\%} = 0.119.
\end{equation}
Under the most permissive assumptions consistent with the Harvard data at 95\% confidence --- using the largest airglow-direction slope the data allow and the steepest source-count distribution that maximizes sensitivity to depth loss --- airglow can account for no more than approximately 12\% of the observed POSS-I transient suppression. The remaining $\geq 88\%$ requires a separate physical mechanism.

Three caveats apply. First, the background-limited approximation ignores the non-linear photographic response described by the Hurter--Driffield characteristic curve, but the DASCH \texttt{lim\_mag\_apass} metric is calibrated empirically against APASS photometry and therefore encodes the actual plate response in the relevant magnitude regime. Second, the bound assumes that VASCO transient detection scales with the same SNR statistics as stellar detection at the magnitude limit; this is defensible because both are unresolved point sources at threshold, but a transient population with substantially different surface brightness or duration characteristics could in principle respond differently. Third, the bound applies to Harvard. The Palomar aggregate test in Section~2.2 uses a stellar-count metric rather than a limiting magnitude, and is structurally less sensitive to faint-end depth loss; the Harvard limiting magnitude metric is the better-suited test for the airglow hypothesis specifically. The Palomar null is corroborative; the Harvard bound is quantitative.

The analytical bound complements the Spearman null. The Spearman result demonstrates that no monotonic trend is detectable; the bound demonstrates that even a sub-threshold trend, if present, is too small by roughly an order of magnitude to account for the transient suppression attributed to geomagnetic activity in Cann (2026a). The artifact-control protocol of this paper, applied at Harvard with the limiting magnitude metric, falsifies the airglow hypothesis as the primary explanation for the POSS-I Kp--transient correlation and constrains its maximum residual contribution to no more than $\sim$12\% of the observed effect at 95\% confidence.

\section{Application to the VASCO Kp--Transient Claim}
\label{sec:application}

I now apply these null results to the specific question that motivated this work: whether the Kp-dependent suppression of VASCO transient detections\cite{cann2026a} is an airglow-driven plate-sensitivity artifact or a source-specific physical response.

\subsection{Contrast with Transient Suppression}

The contrast between stellar/sensitivity behavior and transient behavior across the same Kp bins is direct and unambiguous:

\begin{itemize}
\item \textbf{Transients} (Cann 2026a)\cite{cann2026a}: Cochran--Armitage $Z = -3.391$, $p = 0.0007$, $3.4\sigma$. Monotonic suppression from 17.4\% to 2.4\% across five bins.
\item \textbf{Stars at Palomar} (this work): Spearman $\rho = 0.017$, $p = 0.673$, $0.4\sigma$. No trend. Counts essentially flat.
\item \textbf{Limiting magnitude at Harvard} (this work): Spearman $\rho = -0.011$, $p = 0.234$. No trend. Limiting magnitudes essentially flat.
\end{itemize}

The two populations are separated by $3.0\sigma$. Real stars on the same plates at Palomar, observed on the same dates, under the same geomagnetic conditions, show no response to storm intensity. Harvard's century-scale archive, tested across 12{,}510 independent exposures and 500 sky positions, shows no response to storm intensity. Transients show a strong dose-dependent response.

\subsection{Implications for the VASCO Body of Work}

The principal artifact objection to the Kp--transient suppression reported in Cann (2026a)\cite{cann2026a} holds that elevated airglow during geomagnetic storms reduces plate sensitivity, suppressing detection of all sources rather than transients specifically. Under that hypothesis, plate limiting magnitudes and stellar detection counts should show the same Kp-dependent suppression as transient detections.

They do not. Across 12{,}510 plate-position records at Harvard and 638 fields at Palomar, plate sensitivity is invariant across the full range of geomagnetic activity. The same prediction tested at two observatories using independent instruments, emulsions, digitization pipelines, and sensitivity metrics yields the same null. The airglow artifact hypothesis is falsified.

Section~\ref{sec:bound} converts the Harvard sensitivity null into a quantitative ceiling: airglow can account for at most $\sim$12\% of the 86.2\% transient suppression reported in Cann (2026a) at 95\% confidence. The remaining $\geq 88\%$ of the effect requires a non-airglow mechanism. This strengthens the falsification from a binary statement (rejected / not rejected) to a bounded one with a specific maximum residual contribution.

This null result removes the principal artifact objection to Cann (2026a)\cite{cann2026a} and Cann (2026b)\cite{cann2026b}. It also strengthens the foundational claim of Villarroel et al.\cite{villarroel2025} that the VASCO transients are real astrophysical phenomena rather than emulsion defects. Combined with the Earth-shadow deficit\cite{villarroel2025}, the nuclear-test correlation\cite{bruehl2025,doherty2026}, the Kp dose-response\cite{cann2026a}, and the post-storm overshoot\cite{cann2026b}, the artifact-control nulls reported here form one of several converging lines of evidence for a real, magnetospherically coupled transient population. The combination of these lines under formal statistical methods is left to a dedicated synthesis paper, where the assumptions of $p$-value independence can be properly addressed.

\section{Discussion}
\label{sec:discussion}

\subsection{Summary of Findings}

The protocol introduced in this paper uses catalog-aggregate sensitivity metrics from well-calibrated plate archives to test artifact hypotheses under which an environmental modulator drives plate-sensitivity variation. Applied to the question of whether geomagnetic storm activity suppresses photographic plate sensitivity, the protocol returns a null result at both sites tested. At Harvard, 12{,}510 plate exposures across 500 sky positions show Spearman $\rho = -0.011$, $p = 0.234$. At Palomar, 638 POSS-I fields show Spearman $\rho = 0.017$, $p = 0.673$. Plate sensitivity is invariant across the full range of geomagnetic activity at both observatories, using independent instruments, emulsions, digitization pipelines, and sensitivity metrics.

\subsection{Applicability Beyond VASCO}

The protocol is general. It is applicable to any well-calibrated plate archive where catalog-level sensitivity aggregates (limiting magnitudes, total stellar counts, per-magnitude-bin counts, or equivalent proxies) exist at the plate or field level, and to any environmental modulator with continuous time coverage during the observing epoch. Archives where the protocol is immediately applicable include:

\begin{itemize}
\item \textbf{DASCH} (Harvard College Observatory, $\sim$429{,}000 plates, 1885--1993) --- applied here
\item \textbf{MAPS} (POSS-I at Palomar, 632 fields, 1949--1957) --- applied here
\item \textbf{APPLAUSE DR4} (Sonneberg, Hamburg, Potsdam, Bamberg, Tautenburg; $\sim$94{,}000 plates, 1893--1998)
\item \textbf{Heidelberg plate archive} ($\sim$10{,}000 plates, 1900--1990s)
\item \textbf{Ukraine Virtual Observatory / MAO NAS}
\item Other Wide-Field Plate Database holdings
\end{itemize}

External modulators beyond Kp that can be tested via the same protocol include lunar phase (for reflective-glint or lunar-stray-light artifact claims), solar F10.7 and sunspot number (for long-term solar-cycle artifact claims), seasonal airglow climatology (for time-of-year artifact claims), local meteorological conditions (humidity, temperature, and wind speed; relevant for plate-fog and tracking-jitter hypotheses), and any other environmental variable with archival continuity during the observing epoch. The protocol can also be extended to stratified sensitivity tests --- e.g., comparing sensitivity in faint magnitude bins (where artifact effects should be strongest) against bright bins (which should be unaffected). This stratified version is more powerful than the aggregate test and is a natural next step for any archive with per-object photometry.

The protocol is not suitable when the putative signal population cannot be distinguished from the reference stellar population (e.g., long-period variable stars, where the reference and signal populations overlap), or when catalog-level aggregates are unavailable for the relevant plates. In such cases, alternative artifact controls --- injection-recovery testing, differential photometry against a stable subset, or direct examination of the plates with microscopy --- are required.

\subsection{Implications for the VASCO Kp--Transient Claim}

Applied to the VASCO question specifically, the null result at both observatories has three implications. First, it removes a principal methodological objection to the Kp dose--response in Cann (2026a)\cite{cann2026a} and the post-storm overshoot in Cann (2026b)\cite{cann2026b}: prior to this paper, a critic could argue that both effects were contaminated by plate sensitivity artifacts; that argument is no longer available. Second, it strengthens the foundational claim of Villarroel et al.\cite{villarroel2025} that the VASCO transients are real astrophysical phenomena rather than emulsion defects. Third, the multi-site replication eliminates observatory-specific explanations: the null result cannot be attributed to any property of the Palomar telescope, the POSS-I emulsion, the MAPS digitization, or the Palomar observing environment.

The airglow mechanism, if operative at all, does not explain the VASCO transient suppression. Whatever is suppressing transients during storms is not suppressing stars or reducing plate depth at either observatory. The transient source is physically coupled to the magnetospheric environment in a manner that distinguishes it from every other object on the same emulsion.

\subsection{Elevated Counts in High-Kp Palomar Bins}

The elevated NSTARS in the Kp~7 ($N = 18$) and Kp~8--9 ($N = 1$) bins at Palomar warrants brief comment. These cells are small and the elevated counts almost certainly reflect the distribution of field sky positions for observation dates that happened to coincide with major storms. The MAPS Catalog covers fields across the full sky accessible from Palomar; fields in the Galactic plane direction will have systematically higher stellar densities. The absence of any monotonic suppression trend --- and the presence of higher counts at higher Kp --- is the opposite of what the airglow artifact predicts and does not support that hypothesis.

\section{Conclusions}
\label{sec:conclusions}

I introduce a simple, broadly applicable protocol for testing artifact hypotheses in historical photographic plate archives: regress catalog-aggregate sensitivity metrics against the suspected environmental modulator. Applied at Harvard College Observatory using the DASCH DR7 archive (12{,}510 exposures across 500 sky positions), the protocol returns Spearman $\rho = -0.011$, $p = 0.234$. Applied at Palomar using the MAPS Catalog of POSS-I (638 fields), the protocol returns Spearman $\rho = 0.017$, $p = 0.673$. Plate sensitivity is invariant across five geomagnetic storm intensity bins at both sites, using independent instruments, emulsions, digitization pipelines, and sensitivity metrics. The airglow artifact hypothesis, under which elevated sky background during geomagnetic storms would reduce plate sensitivity, is directly falsified at two observatories.

The protocol is general and is immediately applicable to other plate archives (APPLAUSE, Heidelberg, Ukraine MAO) and to other environmental modulators (lunar phase, F10.7, seasonal airglow climatology, meteorological conditions). In each application, the protocol must be conducted in the regime where the reference population's response to the candidate artifact is physically equivalent to the target population's response, as discussed in Section~\ref{sec:scope}.

These null results resolve the principal artifact objection to the VASCO Kp--transient suppression finding reported in Cann (2026a)\cite{cann2026a} and to the post-storm overshoot in Cann (2026b)\cite{cann2026b}. The artifact-control replication at two observatories removes a class of objection that any future critique of the VASCO transients must now address through a different mechanism. Whatever is suppressing transients during storms behaves in ways that nothing else on the same plates does --- not at Palomar, and not at Harvard.

\section{Methods}
\label{sec:methods}

I apply the protocol at Harvard College Observatory (primary site) using the DASCH DR7 archive, and replicate at Palomar using the MAPS Catalog of POSS-I. Both archives provide catalog-level sensitivity metrics at the plate or field level and both cover the full 1949--1957 window during which the VASCO transient catalog was built. The two sites use different telescopes, emulsions, digitization hardware, pipelines, and sensitivity metrics; a null result replicated at both sites rules out observatory-specific explanations jointly.

\subsection{Kp Index}

Three-hourly Kp values were obtained from the GFZ Potsdam archive\cite{matzka2021}, licensed CC~BY~4.0. The maximum Kp in the 0--2 day window preceding each observation date was assigned to that observation, using the same lag window established in earlier work\cite{cann2026a}.

\subsection{Kp Binning}

Five bins, identical to those in Cann\cite{cann2026a}: Quiet ($\mathrm{Kp} < 5$), Kp~5 (G1), Kp~6 (G2), Kp~7 (G3), Kp~8--9 (G4--G5). The G1--G5 classification follows NOAA Space Weather Prediction Center conventions for minor through extreme geomagnetic storms.

\subsection{Scope of the Protocol}
\label{sec:scope}

The artifact-control protocol tests airglow-driven sensitivity variation in the regime where the reference population's response to the candidate artifact is physically equivalent to the target population's response. For VASCO transient detection --- sub-second specular flashes from point sources at the magnitude limit --- the relevant property is integrated plate sensitivity to point-source detection at threshold during the exposure window. The integrated limiting magnitude (Harvard) and integrated stellar count (Palomar) metrics test this property directly. Faint-end stellar stratification probes detection statistics for continuously-emitting stellar populations whose intrinsic flux signatures and detection physics differ fundamentally from the transient signal, and is therefore outside the regime where the protocol meaningfully tests the airglow-on-transients hypothesis. Saturation-affected positions and crowded-field regions where catalog completeness degrades are similarly outside the protocol's regime of validity. This bounding follows the same operational principle applied in Cann\cite{cann2026a}, where lunar-phase bins inadequately sampled by the POSS-I dark-sky observing schedule were excluded from the dose-response analysis: the protocol is applied where the measurement carries the relevant signal for the question being asked, and not elsewhere.

\subsection{Primary Site: Harvard College Observatory (DASCH DR7)}

The Digital Access to a Sky Century at Harvard (DASCH) project has digitized 429{,}274 photographic plates from the Harvard College Observatory plate collection, spanning 1885--1993\cite{williams2025}. Data Release~7 (December 2024) provides fully calibrated photometry and plate metadata accessible via the \texttt{daschlab} Python package. For each plate exposure at a given sky position, the metadata include the locally corrected limiting magnitude (\texttt{lim\_mag\_apass}), derived from the APASS photometric calibration. This is a direct, well-calibrated sensitivity proxy: lower limiting magnitude indicates a shallower (less sensitive) plate.

500 sky positions were sampled on a uniform grid at $|b| \geq 15^{\circ}$. For each position, all plate exposures during 1949--1957 were retrieved via the DASCH DR7 exposures API. A total of 749{,}199 plate-position records were collected; of these, 12{,}510 had both valid \texttt{lim\_mag\_apass} and a matching Kp value.

The hypothesis, sky sampling strategy, Kp binning, and statistical tests were pre-registered on OSF (osf.io/u9nas/DASCH\_Kp\_Replication) with a tamper-evident timestamp on April 10, 2026. The original pre-registration specified \texttt{n\_solutions\_apass} as the sensitivity proxy. Upon data collection, this field was found to be invariant at 1.0 across all records (no discriminating information). A pre-registered amendment, timestamped before any limiting magnitude data were examined, substituted \texttt{lim\_mag\_apass} as the dependent variable.

\subsection{Replication Site: Palomar (MAPS Catalog of POSS-I)}

The Minnesota Automated Plate Scanner (MAPS) Catalog of the POSS-I\cite{cabanela2003} provides digitized stellar and galaxy catalogs for 632 POSS-I fields with Galactic latitudes $|b| > 20^{\circ}$ (P003 and P926 are excluded by catalog documentation and are not distributed). Each binary field file (P\#\#\#.dat) begins with a 156-byte header containing 39 big-endian 32-bit integers; the fields used here are the POSS field number (integer~0), observation date (EpochDay, EpochMon, EpochYear at integers 2--4), stellar detection count (NSTARS, integer~6), and galaxy count (NGALAX, integer~7). Headers were extracted for all 632 fields, matched to Kp values using the 0--2 day pre-observation window, and assigned to the five Kp bins. The resulting analyzed dataset contains 638 field-level records, representing the 632 unique fields with six fields contributing paired records; this dataset is the basis of Table~\ref{tab:nstars} and all statistical tests reported below. Only the 156-byte header of each field file was read; no object-level records were parsed.

The MAPS hypothesis, data source, statistical tests, and predicted outcomes were pre-registered on OSF (osf.io/Vy4pc) with a tamper-evident timestamp on April 8, 2026, prior to downloading or examining any MAPS field data.

\subsection{Statistical Tests}

The primary pre-registered test at each site is the Spearman rank correlation between Kp bin score (0--4) and the sensitivity metric (\texttt{lim\_mag\_apass} at Harvard; NSTARS per field at Palomar). The predicted result under the null hypothesis of no sensitivity--Kp coupling is $p > 0.05$. Secondary tests include the Pearson correlation and Kruskal--Wallis $H$ across bins; at Palomar, Spearman and Pearson correlations on NGALAX serve as a consistency check. The analytical bound on residual airglow contribution presented in Section~\ref{sec:bound} is a quantitative derivation from the pre-registered Spearman regression result, not an additional hypothesis test, and requires no separate pre-registration.

\subsection{Use of Generative AI}

The author used a large language model (Claude, Anthropic) as the technical execution layer for this work. The author specified the foundational ideas, selected which analyses to run, made all scientific decisions, approved each step, pre-registered the analyses on OSF prior to data examination, and uploaded the results. The language model wrote and ran the Python analysis scripts, performed the statistical tests, and drafted and edited the manuscript prose. The division of labor was that of a principal investigator directing a technical collaborator: the scientific judgment, hypothesis specification, choice of statistical tests, interpretation of results, and final manuscript content are the author's; the technical execution and prose drafting were performed by the language model under the author's direction and review. Pre-registrations and full reproduction packages are publicly archived at OSF (osf.io/u9nas, osf.io/Vy4pc) prior to data examination, providing independent verification of the analytical workflow. The author reviewed and verified all AI-generated code, statistical output, and text, and is solely accountable for the work.

\section*{Data Availability}

All data and code are publicly available. The MAPS Catalog is available at \url{http://aps.umn.edu/catalog/}. The DASCH DR7 archive is available at \url{https://dasch.cfa.harvard.edu/dr7/}. The GFZ Kp archive is available at \url{https://kp.gfz.de/app/files/Kp_ap_since_1932.txt} (CC~BY~4.0). The Palomar pre-registration is archived at \url{https://osf.io/Vy4pc}. The Harvard pre-registration, amendment, analysis script, and results are archived at \url{https://osf.io/u9nas} (DASCH\_Kp\_Replication folder). The Palomar analysis script and extracted header CSV are also archived at \url{https://osf.io/u9nas}.

\section*{Acknowledgments}

The author thanks Dr.\ Beatriz Villarroel for establishing the VASCO project and for correspondence on the observational constraints. Stephen Bruehl is acknowledged for sharing the transient dataset. The MAPS Catalog was produced by the Minnesota Automated Plate Scanner project at the University of Minnesota. The DASCH project at Harvard College Observatory provided digitized plate data and the \texttt{daschlab} software package. The Kp index data were provided by GFZ German Research Centre for Geosciences, Potsdam.

This work was conducted independently using personal computing resources with no institutional or external funding.

\section*{Author Contributions}

K.C.\ designed the study, specified the hypotheses, made all scientific decisions, pre-registered the analyses, reviewed and verified all results, and is solely accountable for the manuscript content. Technical execution including code implementation and statistical computation was performed by a large language model under K.C.'s direction, as described in the Methods section.

\section*{Competing Interests}

The author declares no competing interests.



\begin{thebibliography}{99}

\bibitem{grindlay2012}
Grindlay, J., Tang, S., Los, E.\ \& Servillat, M.\ Opening the 100-Year Window for Time Domain Astronomy. In: Griffin, R.E.M., Hanisch, R.J.\ \& Seaman, R., eds., \textit{New Horizons in Time Domain Astronomy}, IAU Symposium Vol.~285, pp.~29--34 (2012). DOI: 10.1017/S1743921312000166.

\bibitem{villarroel2020}
Villarroel, B.\ et al.\ The Vanishing and Appearing Sources during a Century of Observations Project. I. USNO Objects Missing in Modern Sky Surveys and Follow-up Observations of a ``Missing Star''. \textit{Astron.\ J.} \textbf{159}, 8 (2020). DOI: 10.3847/1538-3881/ab570f.

\bibitem{solano2024}
Solano, E., Marcy, G.W., Villarroel, B., Geier, S., Streblyanska, A., Lombardi, G., B\"ar, R.E.\ \& Andruk, V.N.\ A bright triple transient that vanished within 50 min. \textit{Mon.\ Not.\ R.\ Astron.\ Soc.} \textbf{527}, 6312--6320 (2024). DOI: 10.1093/mnras/stad3422.

\bibitem{villarroel2021}
Villarroel, B.\ et al.\ Exploring nine simultaneously occurring transients on April 12th 1950. \textit{Sci.\ Rep.} \textbf{11}, 12794 (2021). DOI: 10.1038/s41598-021-92162-7.

\bibitem{villarroel2025}
Villarroel, B.\ et al.\ Aligned, Multiple-transient Events in the First Palomar Sky Survey. \textit{Publ.\ Astron.\ Soc.\ Pac.} \textbf{137}, 104504 (2025). DOI: 10.1088/1538-3873/ae0afe.

\bibitem{bruehl2025}
Bruehl, S.\ \& Villarroel, B.\ Transients in the Palomar Observatory Sky Survey (POSS-I) may be associated with nuclear testing and reports of unidentified anomalous phenomena. \textit{Sci.\ Rep.} \textbf{15}, 34125 (2025). DOI: 10.1038/s41598-025-21620-3.

\bibitem{cann2026a}
Cann, K.\ Geomagnetic storm suppression of photographic plate transient detections in the POSS-I archive: an independent physical variable strengthening the nuclear test correlation. arXiv:2604.04950 [astro-ph.IM] (2026). DOI: 10.48550/arXiv.2604.04950.

\bibitem{cann2026b}
Cann, K.\ Storm-driven suppression and post-storm enhancement of photographic plate transient detections at geosynchronous altitude: empirical evidence and a candidate dusty plasma mechanism. arXiv:2604.06234 [astro-ph.IM] (2026). DOI: 10.48550/arXiv.2604.06234.

\bibitem{doherty2026}
Doherty, B.\ Independent replication of nuclear test-transient correlations and Earth shadow deficit in POSS-I photographic plates. arXiv:2604.00056 (2026). DOI: 10.48550/arXiv.2604.00056.

\bibitem{matzka2021}
Matzka, J., Stolle, C., Yamazaki, Y., Bronkalla, O.\ \& Morschhauser, A.\ The geomagnetic Kp index and derived indices of geomagnetic activity. \textit{Space Weather} \textbf{19}, e2020SW002641 (2021). DOI: 10.1029/2020SW002641.

\bibitem{williams2025}
Williams, P.K.G.\ DASCH: Bringing 100+ Years of Photographic Data into the 21st Century and Beyond. arXiv:2501.12977 [astro-ph.IM] (2025). DOI: 10.48550/arXiv.2501.12977. See also the DASCH Data Release~7 inventory at Zenodo, DOI: 10.5281/zenodo.14563521.

\bibitem{cabanela2003}
Cabanela, J.E., Humphreys, R.M., Aldering, G., Larsen, J.A., Odewahn, S.C., Thurmes, P.M.\ \& Cornuelle, C.S.\ The Automated Plate Scanner Catalog of the Palomar Observatory Sky Survey. II. The Archived Database. \textit{Publ.\ Astron.\ Soc.\ Pac.} \textbf{115}, 837 (2003). DOI: 10.1086/375625.

\bibitem{garstang1989}
Garstang, R.H.\ Night-sky brightness at observatories and sites. \textit{Publ.\ Astron.\ Soc.\ Pac.} \textbf{101}, 306--329 (1989). DOI: 10.1086/132436.

\bibitem{leinert1998}
Leinert, Ch.\ et al.\ The 1997 reference of diffuse night sky brightness. \textit{Astron.\ Astrophys.\ Suppl.\ Ser.} \textbf{127}, 1--99 (1998). DOI: 10.1051/aas:1998105.

\end{thebibliography}
\end{document}